%% file: paper.tex
\PassOptionsToPackage{dvipsnames}{xcolor}
\documentclass[9pt,sigconf,screen,authorversion]{acmart}

\usepackage[all]{nowidow}
\usepackage{xcolor}
\usepackage{subcaption}
\usepackage{hyperref}
\usepackage{acronym}

\usepackage[capitalise,noabbrev]{cleveref}
\crefformat{section}{\S#2#1#3}
\crefformat{subsection}{\S#2#1#3}
\crefformat{subsubsection}{\S#2#1#3}
\crefrangeformat{section}{\S#3#1#4 to~\S#5#2#6}
\crefrangeformat{subsection}{\S#3#1#4 to~\S#5#2#6}
\crefrangeformat{subsubsection}{\S#3#1#4 to~\S#5#2#6}
\definecolor{darkblue}{rgb}{0.0, 0.0, 0.55}

\newcommand{\sysname}{\textrm{FaaSMoE}}

\begin{document}

\author{Minghe Wang}
\affiliation{%
    \institution{TU Berlin}
    \city{Berlin}
    \country{Germany}}
\email{mw@3s.tu-berlin.de}
\orcid{0009-0001-3780-5828}

\author{Trever Schirmer}
\affiliation{%
    \institution{TU Berlin}
    \city{Berlin}
    \country{Germany}}
\email{ts@3s.tu-berlin.de}
\orcid{0000-0001-9277-3032}

\author{Mohammadreza Malekabbasi}
\affiliation{%
    \institution{TU Berlin}
    \city{Berlin}
    \country{Germany}}
\email{mm@3s.tu-berlin.de}
\orcid{0009-0002-0627-3600}

\author{David Bermbach}
\affiliation{%
    \institution{TU Berlin}
    \city{Berlin}
    \country{Germany}}
\email{db@3s.tu-berlin.de}
\orcid{0000-0002-7524-3256}

\title{FaaSMoE: A Serverless Framework for Multi-Tenant Mixture-of-Experts Serving}

\acmYear{2026}\copyrightyear{2026}
\setcopyright{cc}
\setcctype[4.0]{by-nc-nd}
\acmConference[MobiSys Workshop '26]{The 24th Annual International Conference on Mobile Systems, Applications and Services}{June 21--25, 2026}{Cambridge, United Kingdom}
\acmBooktitle{The 24th Annual International Conference on Mobile Systems, Applications and Services (MobiSys Workshop '26), June 21--25, 2026, Cambridge, United Kingdom}
\acmDOI{10.1145/3812836.3814785}
\acmISBN{979-8-4007-2712-2/26/06}


\begin{abstract}
Mixture-of-Experts (MoE) models offer high capacity with efficient inference cost by activating a small subset of expert models per input.
However, deploying MoE models requires all experts to reside in memory, creating a gap between the resource used by activated experts and the provisioned resources.
This underutilization is further pronounced in multi-tenant scenarios. 
In this paper, we propose \sysname{}, a multi-tenant MoE serving architecture built on Function-as-a-Service (FaaS) platforms. 
\sysname{} decouples the control and execution planes of MoE by deploying experts as stateless FaaS functions, enabling on-demand and scale-to-zero expert invocation across tenants.
\sysname{} further supports configurable expert granularity within functions, trading off per-expert elasticity for reduced invocation overhead.
We implement a prototype with an open-source edge-oriented FaaS platform and evaluate it using \texttt{Qwen1.5-moe-2.7B} under multi-tenant workloads.
Compared to a full-model baseline, \sysname{} uses less than one third of the resources, demonstrating a practical and resource-efficient path towards scalable MoE serving in a multi-tenant environment.
\end{abstract}

\maketitle

\input{sections/1_introduction.tex}
\input{sections/2_background.tex}
\input{sections/3_architecture.tex}
\input{sections/4_experiment.tex}
\input{sections/6_discussion.tex}
\input{sections/7_conclusion.tex}

\begin{acks}
    Partially funded by the \grantsponsor{BMFTR}{Bundesministerium für Forschung, Technologie und Raumfahrt (BMFTR, German Federal Ministry of Research, Technology and Space)}{https://www.bmftr.bund.de/EN/Home/home_node.html} in the scope of the Software Campus 3.0 (Technische Universit\"at Berlin) program -- \grantnum{BMFTR}{01IS23068}.
\end{acks}

\balance

\bibliographystyle{ACM-Reference-Format}
\bibliography{bibliography.bib}

\end{document}

%% file: sections/1_introduction.tex
\section{Introduction}
\label{sec:introduction} 
Mixture-of-Experts (MoE) models achieve high accuracy while maintaining computational efficiency by activating only a small subset of expert models for each token at run time~\cite{dai2024deepseekmoe,lepikhin2020gshard,fedus2022switch,team2024qwen2,jiang2024mixtral}.
Despite this sparse expert activation, all experts must reside in memory to guarantee coverage, making MoE models resource-intensive to deploy and serve~\cite{fedus2022switch,lepikhin2020gshard}.
This creates a gap between activated and provisioned resources.
Prior work explores expert parallelism, offloading, prediction and deduplication to improve the computation and memory efficiency~\cite{lepikhin2020gshard,rasley2020deepspeed,rajbhandari2021zero,huang2025efficient,zhou2025floe,yu2025fmoe,deshpande2024moesaic}.
However, these approaches primarily optimize per-model efficiency and do not address the residency of inactive experts, causing this resource gap to persist and accumulate across tenants.

In this paper, we present \sysname{}, which uses FaaS as runtime for MoE expert serving to enable multi-tenant expert sharing without manual provisioning. 
We believe that FaaS is well-suited to enable such a shared expert pool because 
(i) its event-driven, scale-to-zero, and stateless execution model naturally aligns with the sparse and on-demand expert activation in MoE architectures, 
and (ii) FaaS platforms inherently provide autoscaling for handling diverse workloads in multi-tenant scenarios~\cite{schirmer2025towards,baldini2017serverless,pfandzelter2020tinyfaas,wang2023lotus}.
In \sysname{}, experts are deployed as stateless FaaS functions and activated on demand.
A core part of \sysname{} is its lightweight control plane that hosts non-expert components of the MoE model, e.g., gating or tokenization, and manages micro-batching, request dispatching, and expert function invocation.
To reduce expert invocation overhead and container replication, \sysname{} allows multiple experts to be grouped within a single FaaS function via configurable expert granularity, introducing a trade-off between invocation efficiency and expert-level elasticity.
We implement a proof-of-concept prototype using the Qwen1.5-MoE-2.7B model running on the open-source FaaS platform tinyFaaS~\cite{pfandzelter2020tinyfaas}.
To evaluate resource efficiency, we compare \sysname{} against a full-model baseline where each tenant hosts a complete model.
Under a multi-tenant workload, \sysname{} reduces average per-tenant CPU usage from 187.81\% to 54.4\%, and memory consumption from 36.25 GB to 12.04 GB, i.e., consumes less than one third of the resources.

To this end, we make the following contributions:
\begin{enumerate}
    \item We present \sysname{}, a FaaS-based MoE serving architecture that deploys experts as stateless FaaS functions, enabling on-demand expert activation and expert pool sharing across tenants~(\cref{sec:architecture}).
    \item We provide a proof-of-concept implementation on an open-source edge-oriented FaaS platform, supporting configurable expert deployment granularity to balance invocation overhead and expert elasticity~(\cref{sec:eva:poc}).
    \item We evaluate the resource efficiency of \sysname{} under multi-tenant workloads using Qwen1.5-MoE-2.7B, comparing various deployment strategies~(\cref{sec:eva:methodology},~\cref{sec:results}).
    \item We provide insights into the system design and design trade-off to guide future distributed multi-tenant MoE serving~(\cref{sec:discussion}).
\end{enumerate}

%% file: sections/2_background.tex
\section{Background and Related Work}
\label{sec:background}
This section provides a brief introduction to MoE model~(\cref{sec:background:moe}), and the FaaS paradigm~(\cref{sec:background:faas}), followed by an overview of related work~(\cref{sec:background:relatedwork}).

\subsection{Mixture-of-Experts Models}
\label{sec:background:moe}
MoE models increase model capacity without increasing compute costs by activating only a small subset of experts for each token~\cite{dai2024deepseekmoe,jiang2024mixtral,team2024qwen2,shazeer2017outrageously}.
Architecturally, MoE models replace dense feed-forward network (FFN) layers with MoE layers. 
Each MoE layer consists of a gating network and a set of experts where the gating network makes routing decisions and selects top-k experts for token processing.
This design enables high expressiveness while activating only a fraction of model parameters.
Prior work has optimized MoE models with, e.g., expert offloading, caching, and token-aware scheduling, to reduce memory pressure and improve throughput~\cite{lepikhin2020gshard,fedus2022switch,rasley2020deepspeed,du2022glam}.
Although optimizing per-tenant model instance, the gap between sparse activation and dense expert residency persists in multi-tenant MoE serving scenarios, and accumulates across tenants.
This makes on-demand expert activation and expert sharing across tenants a necessity.

\subsection{Function-as-a-Service Platforms}
\label{sec:background:faas}
Function-as-a-Service (FaaS) is a cloud execution model where cloud providers handle fine-grained infrastructure management including resource provisioning, auto-scaling, and load balancing, allowing developers to focus solely on function logic~\cite{paper_bermbach2021_cloud_engineering,schirmer2025towards,baldini2017serverless}.
FaaS features event-driven, on-demand, and scale-to-zero invocations, providing fine-grained elasticity with strong cross-tenant isolation~\cite{shahrad2020serverless,schirmer2023nightshift}. 
These characteristics make FaaS well suited for diverse workloads and unpredictable execution patterns, which are common in multi-tenant environments.

\subsection{Related Work}
\label{sec:background:relatedwork}
MoE models have received increasing attention as a promising approach for scaling model capacity while keeping inference cost manageable.
Recent research has explored improvements in monolithic model optimization, multi-tenant model sharing and distributed expert deployment to enhance efficiency and scalability.

Some approaches, e.g.,~\cite{fedus2022switch,lepikhin2020gshard,rajbhandari2022deepspeed}, optimize performance or resource consumption for monolithic MoE deployments.
As such, they can inherently not leverage the access patterns and resulting resource demands of multi-tenant workloads.

There is work focusing expert sharing across model instances.
Deshpande et al.~\cite{deshpande2024moesaic} propose MoESaic to reduce memory footprint by deduplicating identical activated experts across co-located model instances.
However, MoESaic assumes a full model instance per tenant, i.e., inactivated experts of each model instance still reside in memory and consume resources.

There is also work investigating MoE expert distribution across machines or edge devices: 
Xue et al.~\cite{xue2025wdmoe} propose to distribute experts across edge servers and to apply latency-aware expert prediction to minimize transmission latency. 
Liu et al.~\cite{liu2025expert} design a distributed MoE system that deploys experts as persistent microservices using high-performance communication libraries for data center environments. 
Liu et al.~\cite{liu2025optimizing} propose an expert prediction-based serverless MoE serving approach that minimizes the billing cost of a single model instance.
While this approach also use FaaS as runtime, it relies on model and workload specific expert prediction for single-tenant billing optimization.
Overall, these distributed expert deployment approaches primarily focus on communication scheduling, expert placement, or latency optimization within dedicated GPU clusters or serverless settings, and do not target resource efficiency under dynamic and diverse multi-tenant workloads, where expert demand is sparse and unpredictable across tenants.

%% file: sections/3_architecture.tex
\section{Architecture}
\label{sec:architecture}

\begin{figure}
    \centering
    \includegraphics[width=0.95\linewidth]{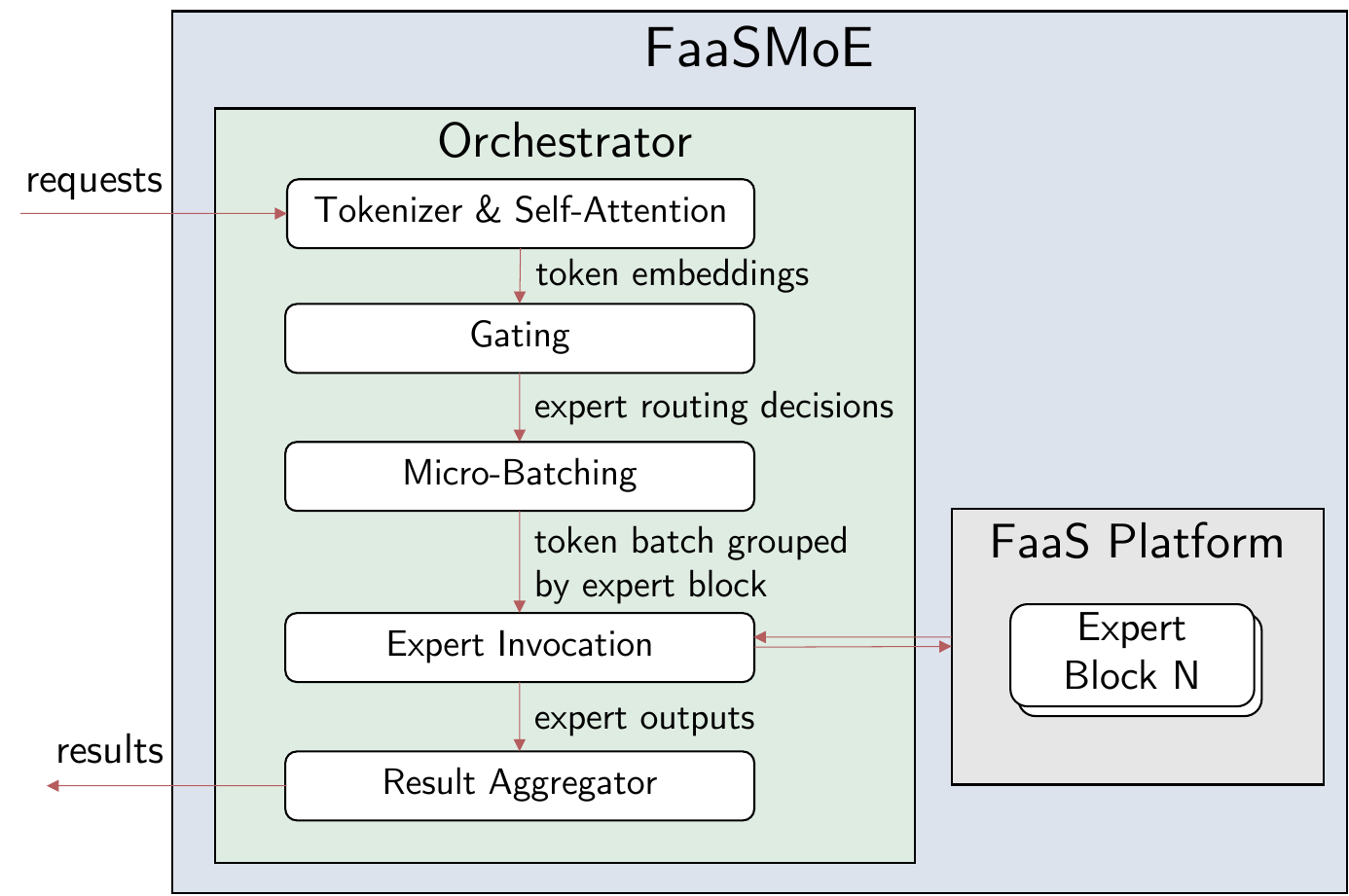}
    \caption{Architecture Overview: \sysname{} decouples MoE inference into a lightweight control plane, i.e., the Orchestrator, and a distributed compute plane comprising MoE experts deployed as stateless FaaS functions.
    }
    \label{fig:architecture}
\end{figure}

MoE experts are stateless by nature, sparsely activated, and independent, yet conventional deployments tie all experts to a monolithic server process. 
This mismatch leads to high memory footprint and limited expert elasticity, especially when different users activate disjoint expert subsets. 
\sysname{} addresses this challenges by decoupling expert execution from the non-expert computation of MoE model and mapping experts to FaaS functions. 
This design treats expert computation as on-demand, elastically scalable functions that can be shared among tenants, while keeping the remainder of the model centrally managed.

As shown in \cref{fig:architecture}, \sysname{} decomposes MoE inference into a lightweight control plane (\emph{Orchestrator}) and a distributed compute plane on the FaaS platform (\emph{Expert Block}). 
The orchestrator consists of a tokenizer, attention layers, and gating modules, which collectively determine the token-to-expert routing pattern.
These components represent a small fraction of an MoE model and incur minimal memory overhead, making them suitable to run using either a centralized orchestrator or scale out with tenants.
After gating, the orchestrator constructs micro-batched routing requests and dispatches them to the corresponding expert functions deployed on the FaaS platform. 
Each expert block contains only MoE experts and performs stateless computation on assigned tokens. 
The FaaS runtime provides isolation, elasticity, and load balancing, enabling frequently used experts to scale out and letting rarely used experts scale to zero. 

A key architectural choice in \sysname{} is the configurable expert granularity per FaaS function. 
Instead of deploying one expert per function, \sysname{} supports packaging multiple experts into an \emph{expert block}, which is exposed as a single FaaS function.
This exposes a tunable trade-off between invocation overhead and expert elasticity, i.e., larger blocks reduce fan-out and invocation overhead and share memory among experts whereas smaller blocks provide finer-grained expert sharing and scaling.
Because the orchestrator retains all non-expert logic, changes to block size do not affect model semantics and routing correctness. 
 
Another aspect of the architecture is the placement flexibility of the orchestrator.
Since the orchestrator holds small size parameters and maintains no cross-request state, it can either be scaled out with tenants or shared. 
Both placements are compatible with the architecture because the orchestrator remains orthogonal to expert execution, i.e., its responsibility for tokenization, gating, routing, and aggregation, is purely structural and does not assume any fixed deployment topology.

Overall, \sysname{} separates immutable model structure of MoE models from elastic expert execution. 
By treating experts as stateless FaaS functions and keeping routing logic lightweight and centralized, this architecture provides resource-efficient multi-tenant sharing, and natural expert elasticity. 
This modular decomposition further supports tuning function granularity, orchestrator placement, and scaling policies based on workload characteristics while preserving correctness and offering deployment flexibility.

%% file: sections/4_experiment.tex
\section{Evaluation}
\label{sec:evaluation}
We evaluate \sysname{} through a proof-of-concept prototype~(\cref{sec:eva:poc}) and experiments under various settings~(\cref{sec:eva:methodology}).

\subsection{Proof-of-Concept Implementation}
\label{sec:eva:poc}
To showcase the feasibility of \sysname{}, we implement a proof-of-concept prototype that we make as an open-source software on GitHub\footnote{\url{https://github.com/Mhwwww/FaaSMoE}}.
The prototype is implemented in Python using Qwen1.5-MoE-2.7B model and tinyFaaS, a lightweight open-source FaaS platform~\cite{pfandzelter2020tinyfaas}.
We extract experts from each MoE layer of the original model and package a configurable number of extracted experts into each FaaS function according to their layer and expert IDs.
During inference, we use asynchronous HTTP calls to invoke experts located on the FaaS platform based on the routing decisions from the gating network.
Specifically, all tokens routed to the same expert block are consolidated into one batched invocation, implementing token-level micro-batching that amortizes serialization and network overhead while preserving expert sparsity and routing correctness.
The asynchronous invocation and the remainder of the model collectively form the orchestrator.
The prototype runs on a CPU-only (no GPU) server with 300 GB memory and disk storage as the backend for the FaaS platform. 

\subsection{Experimental Methodology}
\label{sec:eva:methodology}

To evaluate the system behavior of \sysname{}, we construct a controlled multi-tenant environment with six concurrent clients.
Each client issues five heterogeneous tasks drawn from the BIG-Bench dataset\footnote{\url{http://github.com/google/BIG-bench}}, resulting in thirty requests per experiment run. 
Although the number of requests is small, each request activates multiple experts, resulting in a substantially larger number of expert invocations at the system level.
The experimental setup is designed to compare resource consumption across different deployment strategies to showcase the resource efficiency of expert sharing across tenants. 
Under this setting, the workload is sufficient to expose differences in expert residency, sharing, and orchestration behavior across tenants.

\begin{figure}
    \centering
    \includegraphics[width=0.98\linewidth]{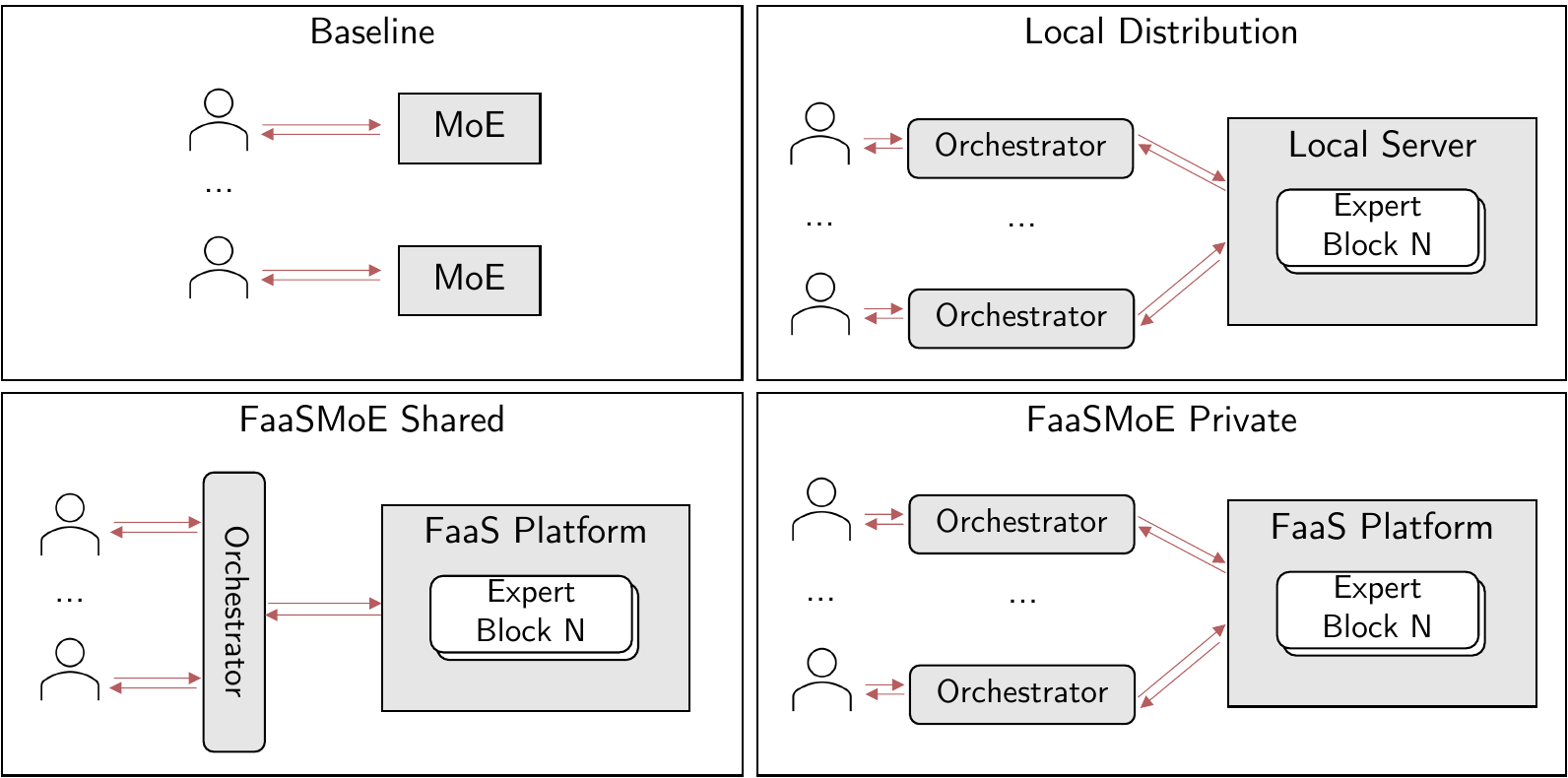}
    \caption{Deployment Strategies: 
    Baseline is the default deployment strategy, where a full MoE model is deployed per tenant. 
    Local Distribution is a non-scaling deployment that separates orchestration from expert execution on a local server.
    \sysname{} Private deploys a per-tenant orchestrator, while \sysname{} Shared further enables cross-tenant orchestrator sharing.
    }
    \label{fig:exp_setups}
\end{figure}

We compare four deployment strategies and show them in \cref{fig:exp_setups}:
\begin{enumerate}
	\item \textbf{Baseline}: Each tenant hosts its own full copy of the MoE model.
	 This represents the default deployment strategy in existing MoE frameworks.
	\item \textbf{Local Distribution}: Each tenant retains only the non-expert components, while all experts are colocated on a shared local server implemented with Uvicorn. 
	    It is a non-scaling setup that serves as a reference to isolate the effect of expert offloading without involving a FaaS platform, which in reality is only a single-tenant option rather than an alternative to \sysname{}.
	\item \textbf{FaaSMoE-Shared}: A single orchestrator instance, running the non-expert modules, gating, batching, and token aggregation, is placed near the FaaS platform, invoking expert-block functions deployed on FaaS.
	\item \textbf{FaaSMoE-Private}: Each tenant hosts its own orchestrator instance, performing routing and scheduling locally before invoking FaaS-based experts. 
\end{enumerate}

Across all setups, we measure CPU utilization in percentage and memory consumption in gigabytes as reported by per-process sampling at 1-second intervals, where 100\% CPU usage represents one fully utilized core.
\emph{Client} denotes the process issuing requests, in the FaaSMoE case, this includes the local orchestrator.
\emph{Server} denotes the expert-hosting backend, i.e., a single Uvicorn server in the Local Distribution strategy, or the FaaS platform for \sysname{}.
Unless otherwise stated, expert functions use a block size of 20 experts per function.

\subsection{Results}
\label{sec:results}
We evaluate the efficiency of \sysname{} by first comparing the four deployment strategies (\cref{sec:eva:results:setups}), followed by analyzing the impact of expert block size (\cref{sec:eva:results:blocksize}).

\subsubsection{Resource Consumption}
\label{sec:eva:results:setups}

\begin{figure}
    \centering
    \subfloat[CPU Usage\label{fig:eva:results:setups_cpu}]{
        \includegraphics[width=0.47\linewidth]{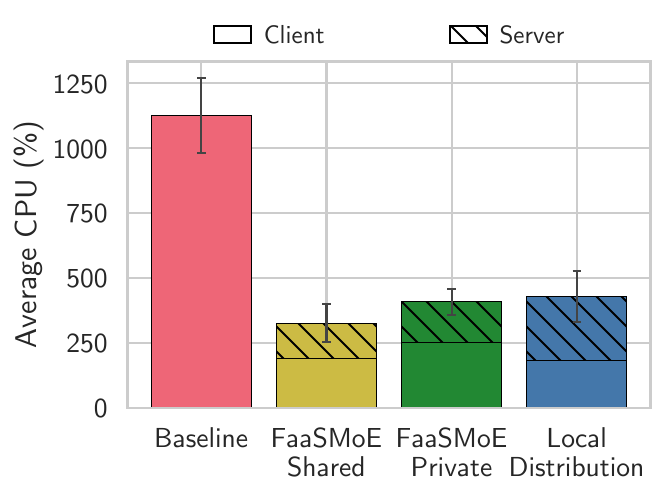}
        }
    \hfill
    \subfloat[Memory Usage\label{fig:eva:results:setups_mem}] {
        \includegraphics[width=0.47\linewidth]{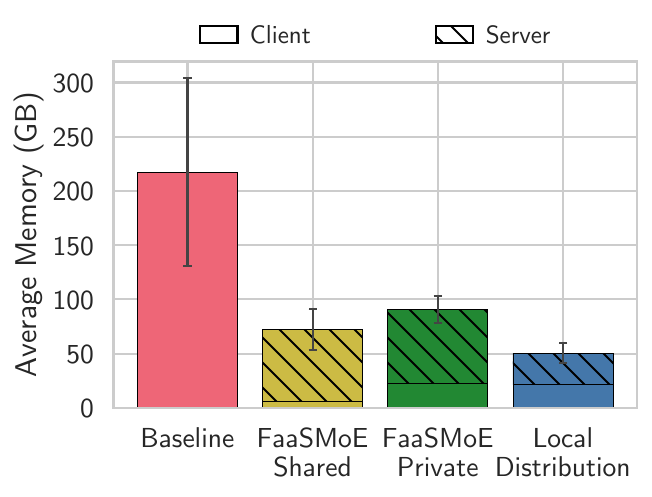}
    }
    \caption{
		Average total CPU and Memory Usage among different experiment settings with expert block size of 20.
        }
    \label{fig:eva:results:setups}
\end{figure}

The results reveal three consistent trends across CPU and memory consumption.
We show different experiment results in \cref{fig:eva:results:setups}, error bars indicate the standard deviation of total consumption.

Baseline exhibits the highest resource footprint.
Hosting a full MoE model at every tenant incurs overhead, the average total CPU usage is 1126.84\%, and average memory usage reaches over 217.52 GB.
This reflects the cost of duplicating all experts per tenant.
\sysname{} shows a similar resource cost and distribution trend on different orchestrator placement.
Both \sysname{} variants reduce total resource consumption compared to baseline while exhibiting different resource distribution pattern depending on orchestrator placement.
FaaSMoE-Shared achieves the lowest average CPU usage at 326.4\% and reduces total average memory to 72.25~GB.
Centralizing the orchestrator enables cross-tenant batching and reduces total request fan-out.
FaaSMoE-Private shows higher average CPU usage at 408.49\% because each tenant maintains its own orchestrator.
This increases serialization and scheduling overhead but improves tenant isolation.
Its memory footprint is also higher at 90.98~GB due to per-orchestrator runtime cost, but still far below the baseline.
Local Distribution removes expert replicas from clients and achieves the lowest total average memory usage at 50.38~GB.
However, all clients share one central expert server with total CPU utilization at 428.67\% on average.
While memory efficient, local distribution lacks the expert elasticity and fine-grained scaling provided by FaaS platforms, leading to a higher CPU usage.
Overall, these results show that a FaaS-based expert pool efficiently shares memory cost of the MoE model among tenants and the orchestrator placement forms a trade-off between batching efficiency and tenant isolation.

\begin{figure}
    \centering
    \subfloat[CPU Usage\label{fig:eva:results:setups_cpu}]{
        \includegraphics[width=0.47\linewidth]{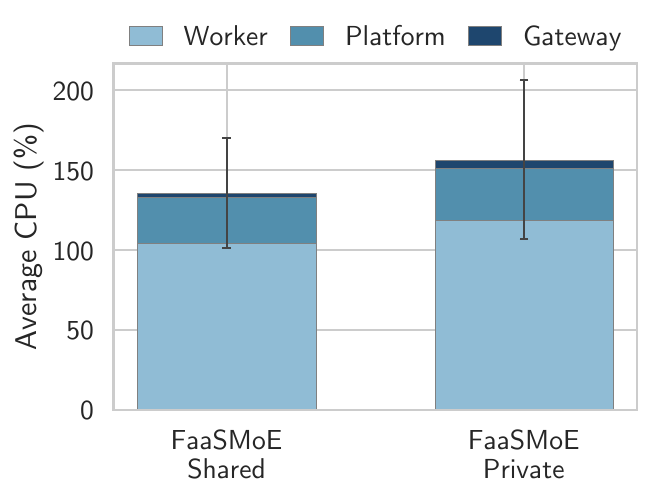}
        }
    \hfill
    \subfloat[Memory Usage\label{fig:eva:results:setups_mem}] {
        \includegraphics[width=0.47\linewidth]{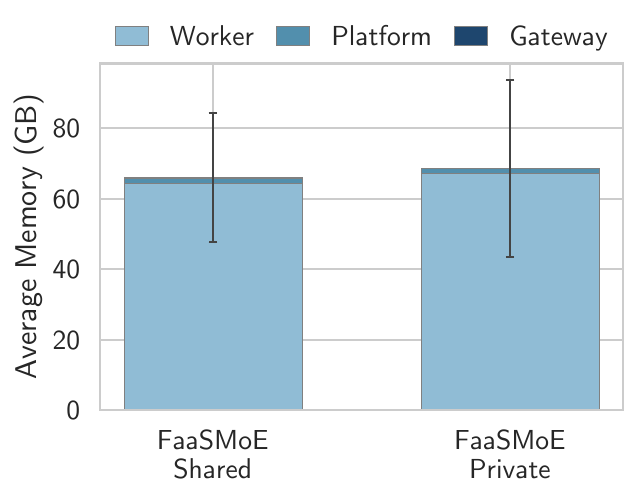}
    }
    
    \caption{FaaS consumption breakdown of \sysname{}: Gateway and platform represent FaaS management consumption and worker represents experts execution.
        }
    \label{fig:eva:results:faas_breakdown}
\end{figure}

To futher understand the overhead introduced by FaaS platforms, we break down the resource consumption of \sysname{} into three components, i.e., expert execution (worker), FaaS platform runtime, and gateway management, as shown in \cref{fig:eva:results:faas_breakdown}.
For both \sysname{} variants, expert execution dominates overall resource usage, while the gateway and platform overheads remain small.
This indicates that the additional overhead introduced by FaaS is modest compared to expert execution.

\subsubsection{Effect of Expert Block Size}
\label{sec:eva:results:blocksize}

\begin{figure}
    \centering
    \subfloat[CPU Usage\label{fig:eva:results:localdist_cpu}]{
        \includegraphics[width=0.48\linewidth]{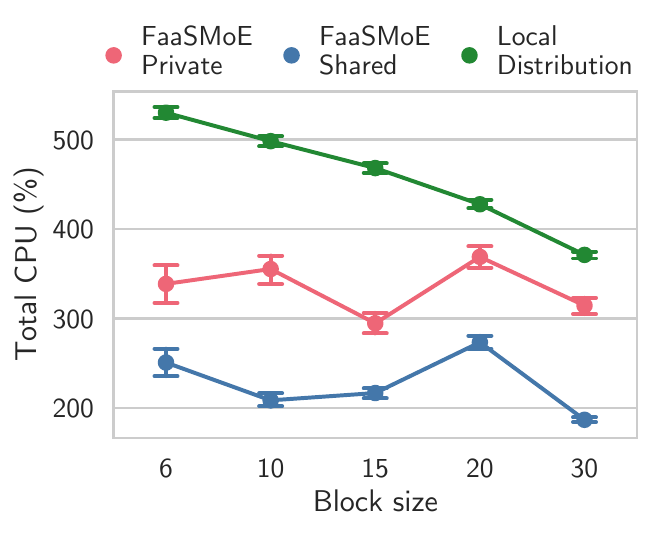}
        }
    \hfill
    \subfloat[Memory Usage\label{fig:eva:results:localdist_mem}] {
        \includegraphics[width=0.48\linewidth]{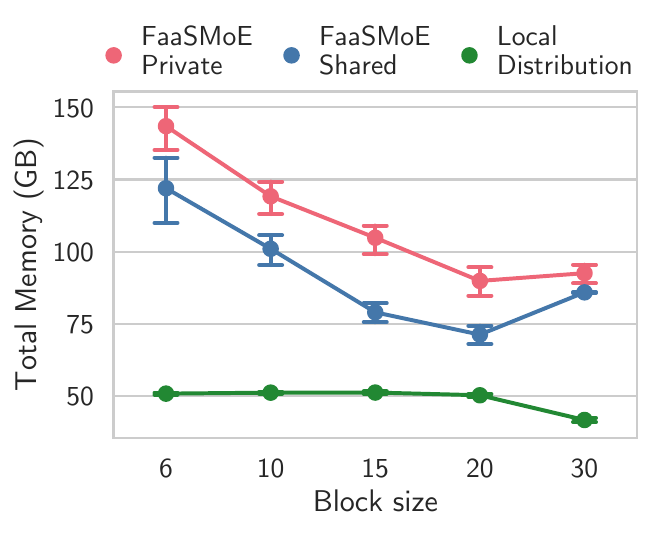}
    }
    
    \caption{
        Average CPU and Memory Usage among different setups with varying block sizes.
        It shows the system overall consumption considering client sending requests and server-side experts processing.
        }
    \label{fig:eva:results:localdist}
\end{figure}

The results in~\cref{sec:eva:results:setups} show that \sysname{} effectively reduces resource consumption through cross-tenant expert sharing. 
There is still a question about what is the optimal number of experts to deploy within a single function to maximize resource efficiency of \sysname{}. 
In this case, we examine how expert block size affects total CPU and memory usage across \sysname{} and Local Distribution and show the results in \cref{fig:eva:results:localdist} with 95\% confidence intervals.

Local Distribution shows a clear monotonic decrease in CPU usage as block size increases, since larger blocks reduce the number of expert groups executed on the centralized server.
In contrast, both \sysname{} variants exhibit non-monotonic CPU trends, i.e., their CPU usage fluctuates across block sizes, with a noticeable peak around block size 20 and a drop at block size 30. 
This reflects the combined effects of batching efficiency, number of FaaS invocations, workload routing overhead rather than a simple linear relationship.
For both \sysname{} setups, memory consumption decreases as block size increases from 6 to 20, reaching the minimum at block size 20, then increases again at block size 30. 
This U-shaped pattern suggests a trade-off between runtime duplication and per-function memory footprint, i.e., overly fine-grained blocks incur runtime overhead, while overly coarse blocks increase resident memory and computation complexity in one function.
Local Distribution, by contrast, maintains almost constant memory usage across block sizes.

%% file: sections/6_discussion.tex
\section{Discussion \& Future Work}
\label{sec:discussion}
\sysname{} explores a resource efficient approach to serving MoE models in multi-tenant environment by decoupling stateless expert execution on FaaS platforms thus enabling on-demand expert invocation and expert sharing across tenants.
The experimental results demonstrate the efficiency of \sysname{} and motivate several discussion points, limitations and directions for future work.

\paragraph{Model Selection}
We evaluate \sysname{} with Qwen1.5-MoE-2.7B model due to its open-source availability and modest resource requirements. 
Qwen1.5-MoE-2.7B also provides fine-grained expert decomposition, with 60 experts and 4 shared experts per MoE layer, which enables exploration of FaaS deployment granularities. 
While MoE models vary in the number of layers and experts, fundamental deployment considerations we study, i.e., expert partitioning, function granularity and orchestrator placement, are properties of the MoE computation pattern rather than the specific backbone.
These observations suggest that the architectural decoupling and expert sharing design of \sysname{} is generalized across a wide range of MoE architectures, which we plan to evaluate in future work.

\paragraph{CPU-only Setup}

Our prototype runs entirely on CPU machines, which aligns with the reality that most mainstream FaaS platforms predominantly support CPU-backed functions, whereas GPU-backed serverless offerings remain limited and costly~\cite{databricks_serverless_gpu,aws_lambda_pricing}. 
Although MoE backbones often benefit from GPU acceleration, the individual experts are typically small MLPs with modest compute requirements.
This makes expert execution a natural fit for CPU-based, short-lived FaaS functions, while obtaining the elasticity and resource isolation benefits provided by FaaS.
By preserving the complete MoE control logic and offloading only expert computation, \sysname{} supports more flexible and accessible deployments beyond GPU-centric infrastructures.

\paragraph{Expert Granularity in FaaS Functions}
\label{para:expert_granularity}
Expert-block granularity, i.e., the number of expert in a single FaaS function, is a key architectural parameter in \sysname{}.
Our evaluation shows that dividing an MoE layer into 3 expert blocks achieves a practical balance, i.e, this reduces invocation frequency comparing to fine-grained designs while still preserving expert elasticity and parallelism absent in coarse-grained ones. 
These results indicate that expert-block granularity should be treated as an architectural design parameter rather than a fixed implementation detail. 
This tunable granularity allows system designers to adjust deployment trade-offs based on cost, latency, and model characteristics. 
Future work can explore adaptive and hot-experts-aware block formation strategies to further optimize runtime efficiency.

\paragraph{Orchestrator Placement and Control-Plane Trade-offs}
The orchestrator serves as the control plane of \sysname{}, and its placement directly affects resource efficiency. 
A centralized orchestrator benefits from cross-tenant micro-batching and connection reuse, reducing CPU consumption.
Distributing orchestrator instances across tenants improves isolation and avoids a single point of failure while incurring per-tenant routing and serialization overhead. 
The placement-sensitive observations highlight a trade-off between global efficiency and per-tenant autonomy, suggesting that hybrid or adaptive placement strategies based on workload characteristics can further improve efficiency under dynamic multi-tenant workloads.

\paragraph{Limitations in Latency and Network}
\sysname{} keeps the original MoE architecture by keeping both experts and control logic unmodified to ensure model accuracy and compatibility.
However, decoupling the expert execution introduces additional network usage and latency due to the frequent communication between orchestrator and remote experts.
Existing work mitigates this by using high-speed RPC and pipelining the overlapping between expert processing and attention computation among batches to hide communication latency~\cite{liu2025expert}. 
This is orthogonally to \sysname{} and can be integrated to enable low-latency communication between client and experts deployed on the FaaS platform.
An alternative for the shared orchestrator instance could be to co-deploy it on the same machine(s) as the FaaS platform, possibly including direct routing to avoid API gateway calls.

%% file: sections/7_conclusion.tex
\section{Conclusion}
\label{sec:conclusion}

MoE models scale model capacity efficiently by activating only a small subset of experts per request but their large memory footprint and fragmented expert usage make them challenging to deploy cost-effectively, particularly for multi-tenant environments.
In this paper, we presented \sysname{}, a serverless-based MoE serving architecture that decomposes experts into stateless FaaS functions and pools them across tenants. 
By leveraging the auto-scaling properties of FaaS platforms, \sysname{} enables on-demand expert activation  and shares memory residency among tenants.
Our evaluation with the Qwen1.5-MoE-2.7B model under multi-tenant workloads shows that comparing to a baseline in which every tenant hosts a complete MoE model, \sysname{} reduces average total memory consumption from 217.52 GB to 72.25 GB, and lowers CPU usage from 1126.84\% to 326.40\%. 
These results suggest that serverless platforms provide an effective runtime environment for MoE inference by enabling elastic, on-demand expert provisioning without requiring per-tenant expert replication, thereby improving resource efficiency and making multi-tenant MoE deployment more practical.